\documentclass{aa}

\usepackage{graphicx}
\usepackage{txfonts}

\newcommand {\mc} {\,$\mu$m}
\newcommand {\cmu} {\,cm$^{-1}$}
\usepackage{graphicx}
\usepackage{txfonts}

\begin{document}

\titlerunning{Processing of ices rich in SO$_2$}
   \title{Thermal and energetic processing of astrophysical ice analogues rich in SO$_2$}

      \author{Z.~Ka\v{n}uchov\'{a} \thanks{corresponding author}
       \inst{1,3}
          \and
          Ph.~Boduch
          \inst{2}
          \and
          A.~Domaracka
          \inst{2}
          \and
          M.E.~Palumbo
         \inst{3}
        \and
        H.~Rothard
           \inst{2}
               \and
           G.~Strazzulla
          \inst{3}
          }
\institute{Astronomical Institute of Slovak Academy of Sciences, SK-05960 Tatransk\'{a} Lomnica, Slovakia\\
         \and
         Centre de Recherche sur les Ions, les Mat\'{e}riaux et la Photonique,
        Normandie Univ, ENSICAEN, UNICAEN, CEA, CNRS, CIMAP, 14000 Caen, France\\
         \and
         INAF-Osservatorio Astrofisico di Catania, via Santa Sofia 78, I-95123 Catania, Italy\\
             }
   \offprints{ Z.~Ka\v{n}uchov\'{a},
   \email{pipovci@gmail.com}}

   \date{Received   / Accepted   }

\abstract
{sulfur is an abundant element in the cosmos and it is thus an important contributor to  astrochemistry in  the  interstellar medium and in the Solar System. Astronomical observations  of the gas and of the solid phases in the dense interstellar/circumstellar regions have evidenced that sulfur is underabundant. The hypothesis to explain such a circumstance is that it is incorporated in some species  in the solid phase (i.e. as frozen gases and/or refractory solids) and/or in the gas phase, which for different reasons have not been observed so far.}
    {Here we wish to give a contribution to the field by studying the chemistry induced by thermal and energetic processing of frozen mixtures of sulfur dioxide (one of the most abundant sulfur-bearing molecules observed so far) and water.}
   {We present the results of a series of laboratory experiments  concerning thermal processing of different  H$_2$O:SO$_2$ mixtures
        and ion bombardment (30\,keV He$^+$) of the same mixtures. We used  in situ Fourier
transform infrared  (FTIR) spectroscopy to investigate the induced effects.}
    {The results indicate that ionic species such as HSO$_{3}^{-}$, HSO$_{4}^{-}$, and S$_2$O$_{5}^{2-}$ are  easily produced. Energetic processing also produces  SO$_3$ polymers and a sulfurous refractory residue.}
   { The produced ionic species exhibit spectral features in a region that, in astronomical spectra of dense molecular clouds, is dominated by  strong silicate absorption. However, such a dominant feature is  associated with some spectral features, some of which have  not yet been identified. We suggest  adding the sulfur-bearing ionic species to the list of candidates to help  explain some of those features. In addition, we suggest that once expelled in the gas phase by sublimation, due to the temperature increase, and/or by non-thermal erosion those species would constitute a class of molecular ions not detected so far.  We also suggest that molecular sulfur-bearing ions could be present on the surfaces and/or in the atmospheres of several objects in the Solar System, for example icy satellites of the giant planets and comets.}


   \authorrunning{Ka\v{n}uchov\'{a} \& al}
  \titlerunning{Thermal and energetic processing of SO$_2$ rich ices}

   \maketitle

\section{Introduction}

Atomic sulfur is an abundant element in the cosmos. Its abundance relative to hydrogen is about 1.32 $\times$ 10$^{-5}$ \citep{Asplund_etal09} and  is thus an important contributor to  chemical evolution in the galaxies and in the Solar System.
However, many astronomical observations (both of the gas and of the solid phases) in the dense interstellar medium (ISM) and in star forming regions have evidenced that sulfur is underabundant, i.e. the sum of sulfur atoms locked in the sulfur-bearing molecules detected so far only accounts for a fraction of its cosmic abundance.

As an example, \cite{Tieftrunk_etal94} summed up the abundances of SO, CS, SO$_2$, and H$_2$S, which are the most abundant S-bearing molecules observed in the gas phase both in low- and high-density molecular clouds. They concluded that these molecules only account  for a fraction of the sulfur abundance in the cosmos, of the order of  10$^{-3}$.
In the solid phase, only OCS \citep{Palumbo_etal95,Palumbo_etal97} and SO$_2$ \citep{Boogert_etal97} have been detected so far in icy grain mantles toward high-mass protostars. Their estimated abundances are low, however, 
and can account for only about 0.5\%, and  0.8-4.0\%, respectively, of the total sulfur abundance \citep{Boogert_etal97,Palumbo_etal97}.

Thus, the problem of the missing sulfur is a hot question in astrochemistry. It is obvious to postulate that sulfur is incorporated in some species -- either in the solid phase and/or in the gas phase -- which for different reasons have not yet been observed. It is important to note that this lack concerns the dense interstellar medium only. In diffuse clouds the amount of gas phase sulfur fully accounts for the total sulfur abundance and rules out the possibility of its depletion  on refractory interstellar grains (see e.g. \citealt{Sofia_etal94}).

In this context some studies have been performed to try to understand which species, although not yet observed in the solid phase, could be present, and then, once expelled to the gas phase by thermal or non-thermal processes, could be searched for in the gas phase. It is  believed that the desorption of grain mantle species into the gas phase, for example after  warming by a protostar or sputtering by energetic cosmic ions, gives an important contribution to the gas phase composition \citep{Modica_Palumbo10,
Palumbo_etal08}. As outlined by \cite{MartinDomenech_etal16} it seems  plausible that an important fraction of the observed S-bearing gas phase species is released from grains because molecules such as H$_2$S, SO$_2$, OCS, SO, H$_2$CS, HCS$^+$, and NS  have abundances that cannot be explained with gas-phase-only chemical models \citep{Doty_etal04, Viti_etal04, Wakelam_etal11,Woods_etal15}.

A molecule that has been considered in particular detail is H$_2$S. This is a very important point because as soon as the medium recondenses (toward the formation of molecular clouds), the high abundance of hydrogen easily produces hydrogenated solid species  such as H$_2$O, CH$_4$, and NH$_3$ which are, indeed, well observed. It is then puzzling  that this is not the case for H$_2$S \citep{Garrod_etal07, Vidal_etal17}. At present it is not clear whether this is due to observational difficulties. The next generation of instruments, namely the James Webb Space Telescope (JWST) could clarify the question.   In the meantime, a number of laboratory experiments simulating the energetic processing of icy mantles on grains in the ISM have demonstrated that energetic processing of solid H$_2$S by ions and photons (UV, X-rays) produce sulfur-sulfur bonds (H$_2$S$_2$ and HS$_2$ are easily formed) and also a refractory polymer-like residue
\citep{Grim_Greenberg87, Moore_etal07, Garozzo_etal10, Jimenez_etal12, Jimenez_etal14}.

It has been  suggested that part of the missing sulfur could be in solid unvolatile refractory grains
\citep{Garozzo_etal10, Jimenez_Munoz11} or it could be released in the gas phase as H$_2$S$_2$,  HS$_2$,  S$_2$ \citep{Jimenez_Munoz11},
 or CS$_2$ \citep[formed when mixtures of H$_2$S and CO are irradiated in the laboratory,][]{Garozzo_etal10}. This would also imply that hydrogen sulfide is not detected since it is easily transformed into different species.

These findings stimulated recent efforts to observe H$_2$S$_2$, HS$_2$, and S$_2$ in the gas phase toward the low-mass warm core IRAS 16293-2422 \citep{MartinDomenech_etal16}. Estimated upper limit abundances of these molecules are up to two orders of magnitude lower than the H$_2$S abundance in the source.   This possibly indicates that gas-phase chemistry after their desorption from the icy mantles efficiently destroys those species.

With the aim of  contributing to the field, we present here the results of a series of experiments conducted at the laboratories of the   Centre de recherche sur les Ions, les MAt\'{e}riaux et la Photonique (CIMAP)-Grand Acc\'{e}l\'{e}rateur National d'Ions Lourds (GANIL)  in Caen (France) and at the  Laboratorio di Astrofisica Sperimentale (LASp) in Catania (Italy). The experiments conducted at CIMAP-GANIL concern the thermal processing of different mixtures H$_2$O:SO$_2$ and the implantation of multicharged sulfur ions in water ice; the experiments conducted in Catania are relative to ion bombardment (30\,keV He$^+$) of the same mixed species. The results indicate that ionic species such as HSO$_{3}^{-}$, HSO$_{4}^{-}$, and S$_2$O$_{5}^{2-}$ are produced by the three processes (thermal, S-implantation in pure water ice, and ion bombardment) and we suggest that they have to be searched for in the inter- and circumstellar regions where they could contribute to the inventory of the missing sulfur atoms.

Our experiments are also relevant to some objects in the Solar System, namely Jupiter's Galilean satellites
\citep[see e.g.][]{Dalton_etal10}. In particular, frozen SO$_2$ is the dominant species at the surface of Io, and it was also observed in cometary comae
\citep[see e.g.][]{Crovisier_Morvan99}. Our results are therefore discussed also in the light of their relevance for these objects.

\section{Experimental procedure}

  The experiments conducted at CIMAP-GANIL concern the thermal processing of different mixtures of H$_2$O:SO$_2$. The frozen samples were prepared by condensing opportune mixtures of water and sulfur dioxide gases on a CsI window at 16\,K.  A fine valve allowed  the deposition rate to be controlled. A nozzle was used to transmit the gas into the high vacuum chamber and onto the cold CsI substrate installed in the  centre of the chamber on a cold finger connected to a closed-cycle helium cryostat. The pressure in the high vacuum chamber was below 10$^{-7}$\,mbar.  The temperature of the substrate was controlled by a carbon resistance and a compound linear thermal sensor (CLTS) situated on the holder, providing a precision of 0.1\,K. After deposition the samples were  heated up at a rate of about 1K/min and IR spectra taken at the chosen temperatures in the spectral range 5000 - 600\,cm$^{-1}$ (2-16.7\,$\mu$m) with a resolution of 1\,cm$^{-1}$. To this end, a Nicolet Magna 550 Fourier Transform Infrared Spectrometer (FTIR) was used. The spectra are taken in transmittance, at normal incidence, and were corrected by a background spectrum recorded before deposition \citep[for more details on the experimental set up, see][]{Ding_etal13}.

In the experiments conducted in Catania,  H$_2$O:SO$_2$ (1:2) mixtures were  accreted onto a cold (16\,K) silicon substrate in a vacuum chamber (P $<$ 10$^{-7}$\,mbar).  Infrared transmittance spectra (resolution of 1\,cm$^{-1}$) were obtained, before and after 30\,keV He$^+$ ion bombardment, by  a Bruker Equinox~55 FTIR spectrometer. Ion beams were produced by an ion implanter (Danfysik 1080-200) and irradiated the sample on a spot greater than the area probed by the infrared beam \citep[for more details on the experimental set up, see][]{Strazzulla_etal01, Allodi_etal13}. As usual in this kind of experiment the molecular ratio of the irradiated mixture is different from that expected in space, which is often dominated by water ice. This is due to the experimental need for  a sufficient number of mother molecules to produce the daughter species.

It is important to note that the stoichiometry of a deposited mixture can be evaluated only approximately. In all of the studied mixtures (i.e. at CIMAP-GANIL and in Catania) we  evaluated the column density of the deposited species (H$_2$O and SO$_2$) from infrared spectroscopy. The results significantly differ from the nominal gas mixtures that we prepared before accretion onto the cold finger. This is due to the different thermodynamic properties of the deposited species.

The IR bands of a given molecule were used  to measure  the  column density $N$ in units of molecules cm$^{-2}$ through the formula
\begin{equation}
N=\frac{\int \tau(\nu)d\nu}{A}
,\end{equation}
where $\tau(\nu)$ is the optical depth (which is 2.3 times the absorbance plotted in the figures) at wavenumber $\nu$ (cm$^{-1}$) and $A$ is the band strength (cm molecule$^{-1}$).

The used band strength values are given in Table~\ref{table_peaks}  together with band peak positions and  assignment. The band strength values are valid for pure species, and  using them to evaluate the column density of each molecule in a mixture introduces a large error that can be as high as 50\%.

\begin{table*}
\centering
      \caption{Peak positions, vibration modes, molecule assignments, and  strength of the bands used to calculate the column density of the deposited H$_2$O and SO$_2$ mixtures. }
         \label{table_peaks}
                                \centering
                                \begin{tabular}{l l c c c l}
            \hline
            \noalign{\smallskip}
\multicolumn{2}{c}{ Peak position} & Vibration & Assignment & Band strength  & Ref. \\
{\cmu}& {\mc}  &  &  & $\times$10$^{-17}$cm~mol$^{-1}$ & \\                             
\noalign{\smallskip}
            \hline
            \noalign{\smallskip}
1149  & 8.703 & ${\nu}_{1}$  & SO$_{2}$    & 0.22  & \citet[][]{Garozzo_etal08} \\
1335  & 7.491 & ${\nu}_{3}$  & SO$_{2}$    & 1.47  & \citet[][]{Garozzo_etal08} \\
1660  & 6.024 & ${\nu}_{2}$     & H$_{2}$O    & 1.2  & \citet[][]{Gerakines_etal95} \\
3280  & 3.045 & ${\nu}_{1}$, ${\nu}_{3}$  & H$_{2}$O & 14 & \citet[][]{Hagen_etal81} \\

\noalign{\smallskip}
\hline
         \end{tabular}
  \end{table*}

\section{Results}

\subsection{Energetic processing}

\begin{figure}
  \centering
\vspace{0cm}
\includegraphics[width=9 cm, angle=0, bb=20 100 600 600]{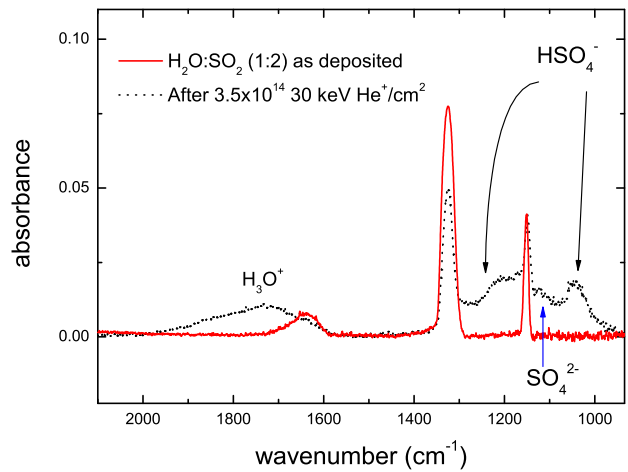}
\caption{Infrared transmittance spectra of H$_2$O:SO$_2$=1:2 ice mixture as deposited at T= 16\,K and after 30\,keV He$^{+}$ ion irradiation (the given fluence corresponds to a deposited dose of 12.3\,eV/16\,u). We note that the broad band around 1200\,cm$^{-1}$ is due to more than one species, including SO$_{3}$ sulfur polymers (see Table~\ref{table_bands}). }
     \label{fig1}
\end{figure}

As an example of the results obtained after ion irradiation of frozen mixtures of water or sulfur dioxide, in  Figure~\ref{fig1} we show the spectra of a deposited
(16\,K) H$_2$O:SO$_2$ (1:2) ice mixture that had a thickness $\sim$0.53\,${\mu}$m, roughly half of the penetration depth of the incoming 30\,keV He$^{+}$ ions calculated by the The Stopping and Range of Ions in Matter (SRIM) software \citep{Ziegler_etal08}. 
Also shown is the spectrum obtained  after irradiation with 3.5$\times$10$^{14}$  30\,keV He$^{+}$/cm$^{2}$. By using the stopping power of the incoming ions as calculated  by the SRIM software \citep{Ziegler_etal08} we find that the  fluence corresponds to a deposited energy (dose) of 12.3\,eV/16\,u \citep[for details see][]{Strazzulla_etal01}.

From Figure~\ref{fig1} it can be easily seen that the intensity of the two SO$_{2}$ bands centred at 1325\,cm$^{-1}$ and 1150\,cm$^{-1}$ diminishes after irradiation. More precisely, the column density of sulfur dioxide decreases from about 5$\times$10$^{17}$\,molecules cm$^{-2}$ to about 3$\times$10$^{17}$\,molecules cm$^{-2}$.  A fraction of SO$_2$ molecules (and of H$_2$O as well) was used to build up new species. In fact, several new bands appeared in the spectrum after ion irradiation (see also Figure~\ref{fig_comparison}). We can observe the formation of the SO$_3$ polymeric chains testified by the presence of the broad band centred at 1200\,cm$^{-1}$ \citep{Moore_etal07}. Features of sulfate SO$_{4}^{2-}$ and bisulfate HSO$_{4}^{-}$ ions are  observed,  as are features  of the counter-ion H$_{3}$O$^{+}$. The results are consistent with the previous findings of \citet{Moore_etal07} who observed the same bands after 800\,keV proton irradiation of H$_2$O:SO$_{2}$ = 3:1 and 30:1 mixtures (at T=86\,K, T=110\,K, and T=132\,K). The peak position of the bands of newly formed sulfur-bearing species observed in the present experiments and in those available in the literature \citep{Moore_etal07} are listed in Table~\ref{table_bands}.

The experiments described so far are relative to ions whose penetration depth is greater than the thickness of the irradiated layers, as  usually occurs with ice mantles on dust grains in inter- and circumstellar environments irradiated by cosmic ions. There are, however, many instances in which the thickness of the irradiated icy layers is much greater than the penetration depth of the ions that remain implanted in the target. This is the case of most of the icy objects in the Solar System  (e.g. satellites of the giant planets, comets, Pluto). Implanted ions --  if they are reactive, like  carbon and sulfur ions -- have the chance  to form molecular species that include the projectile \citep[see e.g.][]{Strazzulla11,Ding_etal13,Lv_etal13}.

Relevant to this paper are the results obtained by \citet{Ding_etal13} concerning  the implantation of S$^{q+}$ (q = 7, 9, 11) ions at an energy range between 35 and 176\,keV in water ice at 80\,K and aimed at simulating  the complexity of the irradiation environment to which the surface of icy satellites of the giant planets, particularly  Europa, are exposed being embedded in the planetary  magnetospheres. The experiments,  performed at the low-energy ion beam facility ARIBE of GANIL in Caen (France), indicate that implantation produces hydrated sulfuric acid  with yields that increase with ion energy. The identification was due to the appearance, in the IR spectra of implanted targets, of a broad feature  characterized by three maxima around 1135\,cm$^{-1}$, 1105\,cm$^{-1}$, and 1070\,cm$^{-1}$ (see the spectrum in the upper panel of Fig.~\ref{fig_comparison}).  Following \citet{Loeffler_etal11} the three observed peaks were assigned to H$_2$SO$_4$, to HSO$_{4}^{-}$ in monohydrate, and to SO$_{4}^{2-}$ in tetrahydrate.

\subsection{Thermal processing}

\begin{figure}[h!]
  \centering
\vspace{0cm}
\resizebox{\hsize}{!}{\includegraphics[width=\textwidth, angle=0]{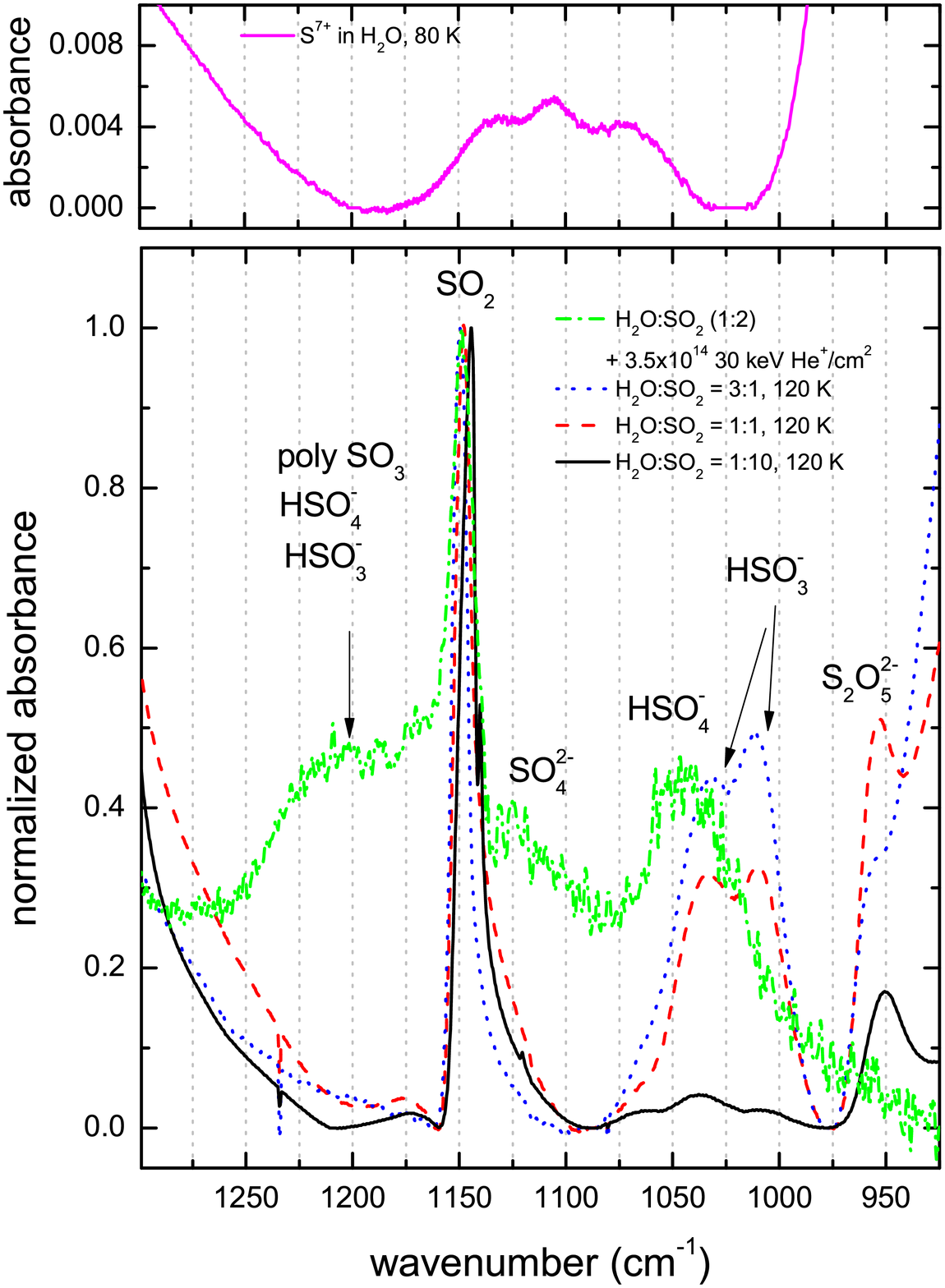}}
\caption{Comparison of the results obtained by three different kinds of ice processing: thermal processing of different H$_2$O:SO$_2$ mixtures, ion bombardment (30\,keV He$^+$) of the same mixed species, and implantation of $S^{7+}$ ions in water ice. }
     \label{fig_comparison}
\end{figure}

\begin{figure}[!ht]
  \centering
\vspace{0cm}
\resizebox{\hsize}{!}{\includegraphics[width=5cm, angle=0]{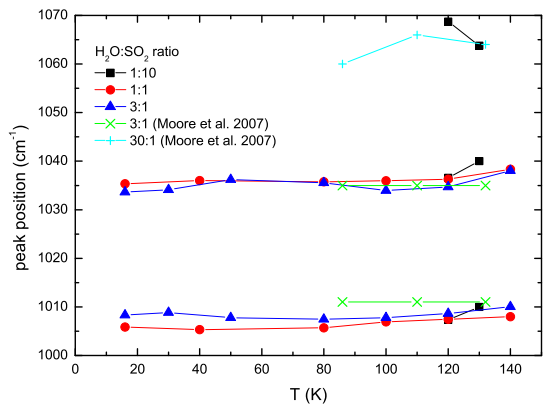}}
\caption{Peak positions  of bisulfite ions (HSO$_{3}^{-}$) in different H$_2$O:SO$_2$ ice mixtures during thermal processing.}
     \label{fig_positionsT}
\end{figure}

\begin{figure}[ht!]
  \centering
\vspace{0cm}
\resizebox{\hsize}{!}{\includegraphics[width=5cm, angle=0]{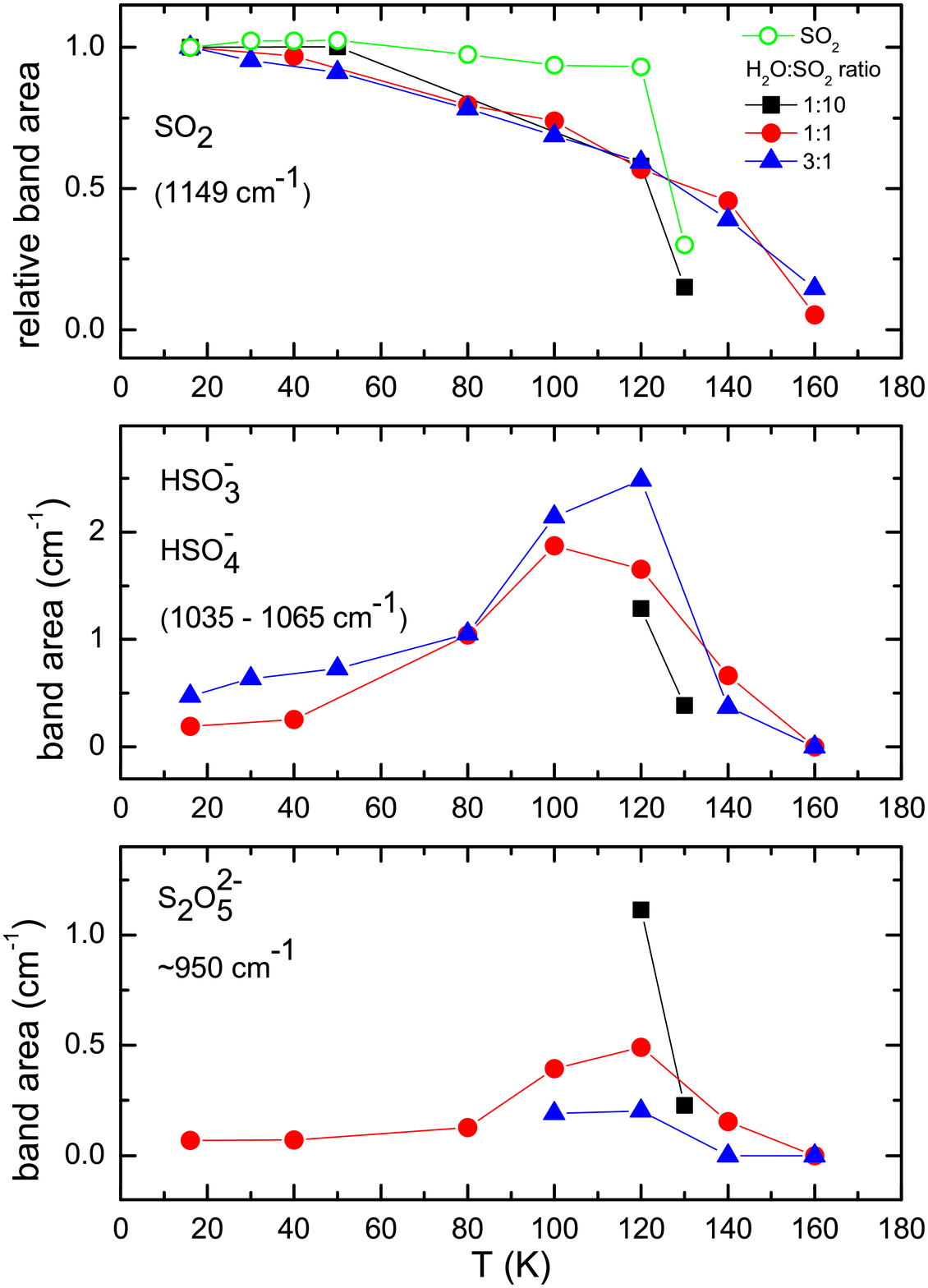}}
\caption{Variations of the relative SO$_{2}$ band area (top panel) and band areas of sulfite and sulfate ions in different H$_2$O:SO$_2$ ice mixtures during thermal processing.}
     \label{fig_relative_bandsT}
\end{figure}

Three samples of icy mixtures with the H$_2$O:SO$_2$ concentration ratios 1:10, 1:1, and 3:1 were deposited at 16\,K and then warmed step by step up to T=160\,K. A blank experiment of pure SO$_2$ was also  performed. Infrared spectra were taken at low temperature and at various steps during the heating of the samples. Spectra taken at 120\,K are plotted in the bottom panel of Fig.~\ref{fig_comparison}. \citet{Moore_etal07}, guided by the works of \citet{Zhang_Ewing02} and \citet{Fink_Sill82}, suggest that the peaks at 1035 and 1011\,cm$^{-1}$ are probably due to the bisulfite ion HSO$_{3}^{-}$ and either one of its reaction products or an isomer.  Another absorbance peak present at around 956\,cm$^{-1}$ (see Fig.~\ref{fig_comparison}) is attributed to S$_{2}$O$_{5}^{2-}$, $meta$-bisulfite \citep{Pichler_etal97, Moore_etal07}. Positions of observed absorption peaks were measured (see Fig.~\ref{fig_positionsT}) and compared with the position of absorption bands identified in the similar heating experiments of H$_2$O:SO$_2$ icy mixtures performed by \citet{Moore_etal07} and later by \citet{LoeHud10} on different ratios of the same mixture. Our finding for 1:1 and 3:1 mixtures are in excellent agreement with the results of \citet{Moore_etal07}.
However, in the experiment of \citet{Moore_etal07} with the mixture 30:1 (i.e. with the lowest SO$_2$ concentration), the absorption feature of the bisulfite ion is located at about 1070-1060\,cm$^{-1}$. We do not observe a peak in this region at low T for a mixture 1:10 (i.e. with the highest SO$_2$ concentration), but only in the spectra taken at high temperature (120\,K and above). This could be explained as being due to the sublimation of SO$_{2}$ and to the drop in its initial high concentration (see Fig.~\ref{fig_relative_bandsT}).

Because we do not know the band strengths of the newly formed bands, we are not able to measure the column density of the species. However, assuming that the band strength values do not depend on the temperature, we can deduce the fate of sulfur dioxide and sulfite ions in the mixtures by measuring  the band areas with increasing temperature.
The band areas of all relevant features and the relative area of the SO$_{2}$ band at 1149\,cm$^{-1}$ were measured and they are plotted in Fig.~\ref{fig_relative_bandsT}. For the newly formed bands, we in fact measured a small initial value  for the band areas (in agreement with the finding by \citealt{Moore_etal07} and \citealt{LoeHud10}) that we attribute to the thermal reactions induced by the water latent heat of condensation of water ice.
When pure SO$_{2}$ ice is heated, its sublimation occurs at 120\,K as evidenced by a drop in the relative absorption band area
(Fig.~\ref{fig_relative_bandsT}). When SO$_{2}$ ice is mixed with water ice,
the area of the 1149\,cm$^{-1}$ band decreases well before 120\,K. Together with the decrease in the SO$_{2}$ band area, the sulfite absorption feature (1035--1065\,cm$^{-1}$) grows with the temperature  up to 120\,K, after which it is lost as the samples are further warmed. Thus, before it sublimates, about 50\% of the sulfur dioxide was used by thermal reactions  with H$_{2}$O for the formation of sulfur-bearing ionic species. At temperatures $\geq$120\,K, the sublimation of SO$_{2}$ takes place and the newly formed species follow the same fate.
The rate (efficiency) of the formation of ionic species slightly depends on the relative concentrations of the two ices in the mixture as shown in the two bottom panels of Fig.~\ref{fig_relative_bandsT}.
We also notice that because of a higher number of water molecules surrounding the SO$_{2}$ molecules, some of these latter are trapped  and sublimate at higher temperatures, as  is commonly observed for many other icy mixtures \citep[e.g.][]{Collings_etal04}.

\subsection{Energetic vs thermal processing}

The spectra of H$_{2}$O:SO$_{2}$ mixtures processed thermally and by ion bombardment, and the spectrum of pure water ice after S$^{7+}$ implantation are  plotted together in  Fig.~\ref{fig_comparison} for comparison. The overview of the bands associated with these molecules for all  analysed experiments is given in Table~\ref{table_bands}.
We note that sulfate and bisulfate ions are the result of radiolytic processes, while bisulfite and meta-bisulfite are produced by thermal processing. This finding can be reciprocally confirmed and explained by the comparison with the results of  non-radiolytic, thermally driven experiments \citep{LoeHud13, LoeHud16}.
\citet{LoeHud13} performed experiments by heating frozen H$_2$O:SO$_2$:H$_2$O$_2$  mixtures, the last component being the main product of water ice radiolysis. They found that sulfate ions are produced when H$_2$O$_2$ is present, in contrast to what happens after thermal processing of the binary mixture  H$_2$O:SO$_2$.
Thus, for the binary mixture we have
\begin{equation}
SO_{2} + H_{2}O{\rightarrow}HSO_{3}^{-}+H^{+}
\label{}
.\end{equation}
By adding H$_2$O$_2$ (product of radiolysis) this is rapidly followed by
\begin{equation}
H_{2}O_{2}+HSO_{3}^{-}+H^{+}{\rightarrow}HSO_{4}^{2-}+H_{3}O^{+}
\label{}
\end{equation}

It is relevant to mention that  \citet{LoeHud13} used these results as a possible  explanation for some of the observations related to the presence and distribution of hydrogen peroxide across Europa's surface and of its lack on Ganymede and Callisto.

Similarly \citet{LoeHud16} studied the thermal processing of the solid mixture H$_2$O:SO$_2$:O$_3$. They demonstrated that thermally driven reactions in solid phase occur below 150\,K, and the main sulfur-bearing species is bisulfate. They also suggested that  SO$_2$ and O$_3$ on the surface of the icy Jovian satellites will efficiently react making detection of these molecules in
the same vicinity unlikely.

\begin{table*}
      \caption{Overview of the position and assignment of the peaks
                        identified in the spectra of various processed mixtures. }
         \label{table_bands}
\begin{center}
                                \begin{tabular}{l l l l}
                                                \hline
            \noalign{\smallskip}
 Peak position & Assignment & Process/Ice & Ref. \\
 {\cmu} &  &  &  \\
\noalign{\smallskip}
            \hline
            \noalign{\smallskip}
   611  & SO$_{4}^{2-}$ & H$^{+}\rightarrow$ H$_{2}$O:SO$_{2}$ (3:1; 30:1) & \citet[][]{Moore_etal07}  \\
         891  & ??? & He$^{+}\rightarrow$ H$_{2}$O:SO$_{2}$ (1:2) &  this work \\
         950  &         S$_{2}$O$_{5}^{2-}$ & thermal, H$_{2}$O:SO$_{2}$ (1:1) & this work   \\
         956  & S$_{2}$O$_{5}^{2-}$     & thermal, H$_{2}$O:SO$_{2}$ (3:1; 30:1) & \citet[][]{Moore_etal07}  \\
         982  & SO$_{4}^{2-}$ & H$^{+}\rightarrow$ H$_{2}$O:SO$_{2}$ (3:1) & \citet[][]{Moore_etal07} \\
         1005 - 1011 & HSO$_{3}^{-}$ & thermal, H$_{2}$O:SO$_{2}$ (3:1; 1:10) & \citet[][]{Moore_etal07}; this work \\
         1035 - 1037 & HSO$_{3}^{-}$ &  thermal, H$_{2}$O:SO$_{2}$ (3:1; 1:1; 1:10) & \citet[][]{Moore_etal07}; this work \\
    1044 & HSO$_{3}^{-}$, HSO$_{4}^{-}$ & He$^{+}\rightarrow$ H$_{2}$O:SO$_{2}$ (1:2) & this work \\
                1052 & HSO$_{4}^{-}$ & H$^{+}\rightarrow$ H$_{2}$O:SO$_{2}$ (3:1; 30:1) & \citet[][]{Moore_etal07} \\
                1060 - 1065 & HSO$_{3}^{-}$ & thermal, H$_{2}$O:SO$_{2}$ (1:10) & this work \\
                1070 & SO$_{4}^{2-}$ &  implantation, S$^{7+}\rightarrow$ H$_{2}$O & \citet[][]{Ding_etal13} \\
                1105 & HSO$_{4}^{-}$ &  implantation, S$^{7+}\rightarrow$ H$_{2}$O & \citet[][]{Ding_etal13} \\
                1110 & SO$_{4}^{2-}$ & H$^{+}\rightarrow$ H$_{2}$O:SO$_{2}$ (3:1; 30:1) & \citet[][]{Moore_etal07} \\
                1120 & SO$_{4}^{2-}$ ? & He$^{+}\rightarrow$ H$_{2}$O:SO$_{2}$ (1:2) & this work \\
                1135 & H$_{2}$SO$_{4}$  &  implantation, S$^{7+}\rightarrow$ H$_{2}$O & \citet[][]{Ding_etal13} \\
                1200 &  poly SO$_3$ & He$^{+}\rightarrow$ H$_{2}$O:SO$_{2}$ (1:2) & this work \\
                1235 & HSO$_{3}^{-}$, HSO$_{4}^{-}$ & H$^{+}\rightarrow$ H$_{2}$O:SO$_{2}$ (3:1; 30:1) & \citet[][]{Moore_etal07}\\
\noalign{\smallskip}
\hline
         \end{tabular}
                                \end{center}
  \end{table*}

 \section{Discussion}

 \subsection{Protostellar regions}

As already said, the only sulfur-bearing molecules observed in the solid phase toward high-mass protostars are OCS 
\citep{Palumbo_etal95,Palumbo_etal97} and SO$_2$ \citep{Boogert_etal97}, but they have low abundances and  can account only for a minor amount of the elemental sulfur. Nevertheless, the presence of SO$_2$ in the icy mantles in protostellar regions has stimulated 
the experiments presented here, aimed at investigating which additional sulfur-bearing species are formed after energetic and/or thermal processing of sulfur dioxide mixed with water ice. Our finding  that ionic species such as HSO$_{3}^{-}$, HSO$_{4}^{-}$, and
S$_2$O$_{5}^{2-}$ are produced raises two questions: (1) are they observed/observable in the interstellar medium in the solid phase and/or in the gas phase after they are desorbed from the icy mantles because of thermal and/or non-thermal mechanisms? and (2) can they give a significant contribution to the inventory of sulfur species?

In order to find an answer to the question of the presence of the ionic sulfur-bearing species in the solid phase, we have given a look at the literature and made a comparison between our experimental results and the observations by \cite{Lacy_etal98} relative to the infrared spectra of four embedded protostars in the 750-1230\,cm$^{-1}$ range. This spectral region is dominated by the very intense silicate band  that complicates the detection of possible further contributing species. However, \cite{Lacy_etal98} were able to detect, for NGC 7538 IRS9, a band at 1110\,cm$^{-1}$ that they attributed to frozen ammonia in a polar water-rich interstellar ice, and several others  near 785, 820, 900, 1030, and 1075\,cm$^{-1}$ that were unidentified. Later, the band at 1030\,cm$^{-1}$ was confirmed by ISO observations and attributed to frozen methanol \citep{Gibb_etal04}. 
Unfortunately, a fitting procedure between astronomical and laboratory spectra is not feasible in the present case. In fact, the profile of the observed features, which  overlap with the dominant silicate band, cannot be well defined. In addition, the relative intensities of the possibly observed features cannot be reproduced by a single laboratory spectrum. Each one should be treated as a single feature, but this would introduce an  indetermination that is too strong. Therefore, here we  can only outline that the sulfur species which we identify in laboratory spectra are in a spectral region where there are several unidentified bands whose peak positions are coincident or very close to those measured in the laboratory. Future astronomical observations (e.g. by the JWST) could help to clarify their attribution.  In particular, it would be interesting to observe MonR2-IRS3 (or similar sources),  a source warm enough to have caused the sublimation of the most volatile species  and retained the less volatile ones \citep{SchuKha03}. 

The lack of an appreciable amount of SO$_2$ in the observed spectra gives some insight into the chemical pathway that drives the formation of sulfur-bearing species on icy mantles. It is in fact thought that sulfur atoms accreting on grains are mostly hydrogenated and should produce H$_2$S; however, this is not observed. As suggested by \citet{Garozzo_etal10} this could be due to energetic processing by cosmic ions that in presence of oxygen and carbon bearing species (e.g. CO) easily converts hydrogenated sulfur into other species, including SO$_2$. However,  the further ion processing of SO$_2$ in the presence of oxygen and carbon bearing species reduces the amount of SO$_2$ in favour of other species such as those  investigated here. In addition, the lack of  SO$_2$ after sulfur implantation into water ice has already been evidenced by \citet{Ding_etal13}. Those authors outlined that  SO$_{2}$ is formed by the addition of S to O$_{2}$ (and/or HO$_{2}$), but it is easily converted to hydrated sulfuric acid  via

\begin{equation}
SO_{2} + H_{2}O_{2}{\rightarrow}H_{2}SO_{4}
\label{}
\end{equation}

In other words, atomic sulfur that hits a grain surface is efficiently converted by energetic and thermal processes to more complex sulfur-bearing species rather than being accumulated as sulfur dioxide. In this scenario, we can suggest that these species contribute to the inventory of sulfur-bearing species, but at present it is not possible to establish whether their abundance can significantly contribute to solving the question of the missing sulfur because we are not able, due to the lack of suitable band strength values and the paucity of observational evidences, to measure the column density of these species in the solid phase. In addition, these ionic species and the fragments of SO$_3$ polymers have not yet been observed in the gas phase after their sublimation when the temperature increases, for example  in the regions nearer to the forming star.
We hope that the results presented here  stimulate the search for these species in the gas phase in opportune environments by using  the ALMA facility, for example. Adequate laboratory studies on the relevant spectral line parameters of the sulfur-bearing species are also necessary.

\subsection{Solar System objects}

The experimental results presented here are relevant to a number of objects in the Solar System where sulfur dioxide has been observed or  is presumed to be present. These objects include Jupiter's Galilean satellites Io, Europa, Ganymede, and Callisto. Being embedded in the Jovian magnetosphere, they are exposed to the complex flux of low- (plasma) and high-energy electron and ion bombardment \citep{Dalton_etal10}.

Io's surface is in fact dominated by sulfur dioxide, which is expelled from the very intense volcanic activity triggered  by tidal effects \citep{Peale_etal79}. Although it is thought that Io has lost nearly all of its hydrogen \citep{ZoloFeg99}, the detection of hydrogen pickup ions by Galileo's plasma analyser \citep{FrankPat99} in the space surrounding the satellite raised the question regarding its origin. A first suggestion was hydrogen sulfide \citep{NashHow89}. However, the abundance of this compound is very low (with an upper limit of 10$^{-4}$) with respect to SO$_{2}$ \citep{SchmiRodr03}. The next candidates as hydrogen bearing species are then water ice and/or  hydrate materials whose absorption bands around 3150\,cm$^{-1}$  were tentatively observed \citep{Salama_etal94, Carlson_etal97}. It is, however, also possible that the detected flux of hydrogen ions comes from the Jovian magnetosphere and not from the satellite.

In this scenario, and assuming that hydrogen bearing species are present on Io's surface along with the certain presence of intense fluxes of energetic ions and electrons, radiolytic products are likely to be present. The most abundant should be de-hydrated species such as the fragments of the elemental sulfur residue formed after ion bombardment of pure sulfur dioxide \citep{GomisStraz08}. The sulfur residue could be responsible of the observed red slope in the near-infrared/visible spectral region of Io's spectra and of the molecular fragments S$_{4}$ and S$_{8}$ \citep[see e.g. Fig. 4 in ][]{Dalton_etal10}. Much less abundant (and difficult to observe) are the hydrated ions that we have synthesized in the experiments described here; nevertheless,  they merit further investigation. 

The present experiments are of primary relevance to the remaining three water ice dominated Galilean satellites and for the other icy satellites orbiting Jupiter, Saturn (e.g. Enceladus), and Uranus. As already mentioned above,  \citet{Ding_etal13} have demonstrated 
experimentaly that magnetospheric sulfur ions implanted in  Europa's surface produce hydrated sulfuric acid. The production rate is high enough to explain the quantity of hydrated sulfuric acid on the surface of Europa inferred to be present by modelling the near infrared (2 \mc) water ice band \citep{Dalton_etal13} as observed by Galileo the Near-Infrared Mapping Spectrometer (NIMS). However, the bands due to hydrated sulfuric acid in the laboratory spectra of sulfur-implanted water ice targets  appear in the 1100\,cm$^{-1}$ region (see Fig.~\ref{fig_comparison} and Table~\ref{table_bands}). Such a spectral region has been investigated by instruments on board  Voyager and Cassini  and will be investigated by  instruments on board  the James Webb Space Telescope (JWST).  The data collected by flyby and orbiter missions is incomplete and/or collected under imperfect illumination conditions (e.g. high phase angles). Therefore, the contribution of JWST will be relevant and we suggest that a particular effort should be made to identify the hydrated sulfuric acid features, particularly on Europa which exhibits surface regions exposed to very intense fluxes of energetic sulfur ions \citep{Dalton_etal10}. sulfur-bearing ions should also  be searched for in the exospheres of the icy satellites where they could be expelled by thermal and  non-thermal processes.

Our experiments are also relevant to comets and to all of the small objects in the outer Solar System
(trans-Neptunian objects).   Several sulfur-bearing species have already been observed in different families of comets \citep[see e.g.][]{Crovisier_Morvan99, Crovisier06}.

The question of the type and abundance of  sulfur-bearing species was revised by \citet{Calmonte_etal16} based on the results obtained by the Rosetta Orbiter Spectrometer for Ion and Neutral Analysis/Double Focusing Mass
Spectrometer  in the coma of comet 67P/Churyumov-Gerasimenko. Those authors measured the abundances of the species that were previously known to be present on comets, namely H$_{2}$S, OCS, SO, S$_{2}$, SO$_{2}$, and CS$_{2}$. \citet{Calmonte_etal16} detected   S$_{3}$, S$_{4}$, CH$_{3}$SH, and C$_{2}$H$_{6}$S for the first time, and they concluded that the derived total elemental sulfur abundance of 67P does not show any sulfur depletion. In  addition, those authors presented results indicating that sulfur-bearing species have been processed by radiolysis in the pre-solar cloud and that at least some of the ice from this cloud has survived in comets up to the
present. This conclusion is in fact based on experimental results that show how ion irradiation of sulfur-bearing  species produce a  solid unvolatile sulfur-rich residue \citep{GomisStraz08,Woods_etal15} and also molecules originally not present such as CS$_2$ \citep[formed when mixtures H$_2$S and CO are irradiated in the laboratory,][]{Garozzo_etal10} and the ionic species discussed here that we suggest should be searched for.

\section{Conclusion}

In this paper we have described the results of a series of experiments concerning thermal and energetic processing of SO$_2$ ices mixed with water ice. The  results indicate  that ionic species  such as HSO$_{3}^{-}$, HSO$_{4}^{-}$, and S$_2$O$_{5}^{2-}$ are formed. Ion bombardment also produces SO$_3$ polymers and a sulfur-rich refractory residue. The results have been discussed in view of their potential relevance to the debate on the missing sulfur in the interstellar and circumstellar regions, and to the chemical evolution of the surfaces of icy objects in the Solar System. The results can be summarized as follows:
\\
-- We suggest that sulfur-bearing ionic species could be synthesized on interstellar icy grain mantles by energetic or thermal processes. These species and the fragments of SO$_3$ polymers sublimate when the temperature increases, for example in the regions close to the forming star. This finding should stimulate theoretical, observational, and experimental researches. It is in fact important to theoretically investigate the contribution of the ionic species expelled in the gas phase to the chemistry of those regions. At the same time experimental efforts to measure the rotational spectra of these molecules are necessary in order to allow  the observers to identify them through astronomical observations. \\
-- Hydrated sulfuric acid formed after sulfur ion implantation in water ice produces alterations of the shape of its 2\,$\mu$m band as already observed \citep{Dalton_etal10}. It also gives origin to a multi-peaked band in the mid-IR spectral region \citep{Ding_etal13}, which  should be searched for in the spectra of water-dominated solid surfaces. sulfur-bearing ionic species should be searched for in the gas phase after being released from the surface by thermal and/or non-thermal processes. \\
- On Io it is possible that non-hydrated sulfur-bearing species  have  already been observed. Only a small amount of hydrated sulfur-bearing species is predicted to be present on Io's surface, if any. \\
- The inventory of sulfur-bearing molecules recently implemented  by the Rosetta finding \citep{Calmonte_etal16} supports evidence for the occurrence of radiolysis. If so, that inventory would be even more implemented including the species discussed here.\\

Considering the ensemble of the results presented here, and in agreement with recent findings by the NASA-Goddard group (e.g. \citealt{LoeHud16}) the present  work suggests that the combined effect of thermal and radiolytic processes  plays a fundamental role in driving the chemical evolution of sulfur-bearing ices.

\begin{acknowledgements}

This work was supported by the  Italian Ministero dell'Istruzione, dell'Universit\`{a} e della Ricerca through the grant Progetti Premiali 2012-iALMA (CUP C52I13000140001). G.S. was supported by the Italian Space Agency (ASI 2013-056 JUICE Partecipazione Italiana alla fase A/B1) and by the European COST Action CM1401-Our Astrochemical History. Z.K. was supported by VEGA - The Slovak Agency for Science, Grant No. 2/0032/14. This work was also supported by COST Action TD1308 - ORIGINS.

\end{acknowledgements}


\begin{thebibliography}{}


\bibitem[\protect\citeauthoryear{Allodi et al.}{2013}]{Allodi_etal13} Allodi, M.~A., Baragiola, R.~A., Baratta, G.~A., et al. 2013, Space Sci. Rev., 180, 101


\bibitem[\protect\citeauthoryear{Asplund et al.}{2009}]{Asplund_etal09} {Asplund, M., Grevesse, N.,  Sauval, A. J., \& Scott, P. 2009, ARAA. 47, 481}

\bibitem[\protect\citeauthoryear{Boogert et al.}{1997}]{Boogert_etal97} Boogert, A. C. A., Schutte, W. A., Helmich, F. P., Tielens, A. G. G. M., \& Wooden, D. H. 1997, A\&A, 317, 929

\bibitem[\protect\citeauthoryear{Calmonte et al.}{2016}]{Calmonte_etal16} Calmonte, U., Altwegg, K., Balsiger, H. et al. 2016, MNRAS 462, S253

\bibitem[\protect\citeauthoryear{Carlson et al.}{1997}]{Carlson_etal97} Carlson, R. W., Smythe, W. D., Lopes-Gautier, R. M. C., et al., 1997, GRL 24, 2479

\bibitem[\protect\citeauthoryear{Collings et al.}{2004}]{Collings_etal04}
    Collings, M. P., Anderson, M. A., Chen, et al., 2004, MNRAS 354, 1133

\bibitem[\protect\citeauthoryear{Crovisier}{2006}]{Crovisier06} Crovisier J. 2006, MolPhys, 104, 2737

\bibitem[\protect\citeauthoryear{Crovisier and Bockel\'{e}e-Morvan}{1999}]{Crovisier_Morvan99} Crovisier, J. \& Bockel\'{e}e-Morvan, D., 1999.  SpSciRev 90, 19

\bibitem[\protect\citeauthoryear{Dalton et al.}{2010}]{Dalton_etal10} Dalton, J. B.,  Cruikshank, D. P., Stephan, K., et al. 2010, SSRv, 153, 113

\bibitem[\protect\citeauthoryear{Dalton et al.}{2013}]{Dalton_etal13} Dalton, J. B., Cassidy, T., Paranicas, C., Shirley, J. H., Prockter, L. M., Kamp, L. W., 2013, PlSpSci 77, 45

\bibitem[\protect\citeauthoryear{Ding et al.}{2013}]{Ding_etal13} Ding, J. J., Boduch, P., Domaracka, A., Langlinay, T., Lv, X. Y., Palumbo, M. E., Rothard, H., Strazzulla, G., 2013, Icarus 226, 860

\bibitem[\protect\citeauthoryear{Doty et al.}{2004}]{Doty_etal04} Doty, S.~D., van Dishoeck, E.~F., \& Tan, J. 2004, BAAS, 36, 1505

    \bibitem[\protect\citeauthoryear{Ferrante et al.}{2008}]{Ferrante_etal08}Ferrante, R. ~ F., Moore, M. ~H., Spiliotis, M. ~M., \& Hudson, R. ~L. 2008, ApJ 684,1210

\bibitem[\protect\citeauthoryear{Fink and Sill}{1982}]{Fink_Sill82} Fink, U., Sill, G.~T. 1982, The infrared spectral properties of frozen volatiles. In: Wilkening, L. (Ed.), Comets. Univ. of Arizona Press, Tucson, pp. 164--202

\bibitem[\protect\citeauthoryear{Frank and Paterson}{1999}]{FrankPat99} Frank, L.A., \&  Paterson, W.R., 1999, GRL 104, 21345

\bibitem[\protect\citeauthoryear{Garrod et al.}{2007}]{Garrod_etal07} Garrod, R.~T., Wakelam, V. \& Herbst, E. 2007, A\&A 467, 1103

\bibitem[\protect\citeauthoryear{Garozzo et al.}{2008}]{Garozzo_etal08} Garozzo M., Fulvio, D., Gomis, O., Palumbo M.~E., Strazzulla G. 2008, Planet. Space Sci., 56, 1300--1308

\bibitem[\protect\citeauthoryear{Garozzo et al.}{2010}]{Garozzo_etal10} Garozzo M., Fulvio D., Ka\v{n}uchov\'{a} Z., Palumbo M.~E., Strazzulla G. 2010, A\&A, 509, A67

    \bibitem[\protect\citeauthoryear{Gerakines et al.}{1995}]{Gerakines_etal95} Gerakines, P.~A., Schutte, W.~A., Greenberg, J.~M., van Dishoeck, E.~F. 1995, A\&A, 296, 810


\bibitem[\protect\citeauthoryear{Gibb et al.}{2004}]{Gibb_etal04} Gibb, E. ~ L., Whittet, D. ~ C. ~ B., Boogert, A. ~ C. ~ A., \& Tielens, A.~G.~G.\`M. 2004, ApJS 151,35

\bibitem[\protect\citeauthoryear{Gomis and Strazzulla}{2008}]{GomisStraz08} Gomis, O., \& Strazzulla, G., 2008,  Icarus 194, 146

\bibitem[\protect\citeauthoryear{Grim and Greenberg}{1987}]{Grim_Greenberg87} Grim, R.~J.~A., \&  Greenberg, J.~M. 1987,  A\&A, 181, 155

\bibitem[\protect\citeauthoryear{Hagen et al.}{1981}]{Hagen_etal81}Hagen, W., Tielens, A.~G.~G.~M., Greenberg, J. ~ M., 1981, ChemPhys 56, 367

\bibitem[\protect\citeauthoryear{Jim\'{e}nez-Escobar and Mu\~noz Caro}{2011}]{Jimenez_Munoz11} Jim\'{e}nez-Escobar A., \& Mu\~noz Caro, G. M., 2011, A\&A, 536, A91

\bibitem[\protect\citeauthoryear{Jim\'{e}nez-Escobar et al.}{2012}]{Jimenez_etal12} Jim\'{e}nez-Escobar A., Mu\~noz Caro, G. M., Cicarelli, A., et al. 2012, ApJ, 751, 343

\bibitem[\protect\citeauthoryear{Jim\'{e}nez-Escobar et al.}{2014}]{Jimenez_etal14} Jim\'{e}nez-Escobar A., Mu\~noz Caro, G. M., \& Chen, Y.-J. 2014, MNRAS, 443, 343

\bibitem[\protect\citeauthoryear{Lacy et al.}{1998}]{Lacy_etal98} Lacy, J. H., Faraji, H., Sandford, S. A., \& Allamandola, L. J., 1998, ApJ 501, L105

\bibitem[\protect\citeauthoryear{Loeffler et al.}{2010}]{LoeHud10} Loeffler, M.~J., \& Hudson, R.~L., 2010, GRL  37, L19201
\bibitem[\protect\citeauthoryear{Loeffler et al.}{2013}]{LoeHud13} Loeffler, M.~J., \& Hudson, R.~L., 2013, Icarus 224, 257
\bibitem[\protect\citeauthoryear{Loeffler et al.}{2016}]{LoeHud16} Loeffler, M.~J., \& Hudson, R.~L., 2016, ApJL 833, L9
\bibitem[\protect\citeauthoryear{Loeffler et al.}{2011}]{Loeffler_etal11} Loeffler, M.~J., Hudson, R.~L., Moore, M.~H., \& Carlson, R.~W., 2011, Icarus 215, 270

  \bibitem[\protect\citeauthoryear{Lv et al.}{2013}]{Lv_etal13} Lv, X. Y.,  Boduch, P., Ding, J. J., Domaracka, A., Langlinay, T., Palumbo, M. E., Rothard, H., Strazzulla, G., 2014, MNRAS 438, 922

\bibitem[\protect\citeauthoryear{Mart\'{i}n-Dom\'{e}nech et al.}{2016}]{MartinDomenech_etal16}
Mart\'{i}n-Dom\'{e}nech R., Jim\'{e}nez-Serra I., Munoz Caro G. M.,  M\"{u}ller H. S. P.,  Occhiogrosso A., Testi L.,  Woods P. M., \& Viti S., 2016, A\&A 585, A112

\bibitem[\protect\citeauthoryear{Modica and Palumbo}{2010}]{Modica_Palumbo10} Modica, P. \& Palumbo, M.~E. 2010, A\&A 519, A22

\bibitem[\protect\citeauthoryear{Moore et al.}{2007}]{Moore_etal07}Moore, M.~H., Hudson, R.~L., \& Carlson, R.~W. 2007, Icarus, 189, 409

\bibitem[\protect\citeauthoryear{Nash and Howell}{1989}]{NashHow89}  Nash, D.B., \& Howell,  R.R., 1989, Sci 244, 454

\bibitem[\protect\citeauthoryear{Palumbo et al.}{1995}]{Palumbo_etal95}Palumbo, M.~E., Tielens, A.~G.~G.~M., \& Tokunaga, A.~T. 1995, ApJ, 449, 674

\bibitem[\protect\citeauthoryear{Palumbo et al.}{1997}]{Palumbo_etal97}Palumbo, M.~E., Geballe, T.~R., \& Tielens, A.~G.~G.~M. 1997, ApJ, 479, 839

\bibitem[\protect\citeauthoryear{Palumbo et al.}{2008}]{Palumbo_etal08} Palumbo, M.~E., Leto, P., Siringo, C. \& Trigilio, C. 2008, ApJ 685, 1033

\bibitem[\protect\citeauthoryear{Peale et al.}{1979}]{Peale_etal79}   Peale, S.J., Cassen, P., \&  Reynolds, R.T., 1979, Sci 203, 892

{\bibitem[\protect\citeauthoryear{Pichler et al.}{1997}]{Pichler_etal97} Pichler, A., Fleissner, G., Hallbrucker, A., Mayer, E.,
1997, J. Mol. Struct. 408, 521}

\bibitem[\protect\citeauthoryear{Salama et al.}{1984}]{Salama_etal94} Salama, F., Allamandola, L. J., Sandford, S. A., et al., 1994, Icarus 107, 413

\bibitem[\protect\citeauthoryear{Schmitt and Rodriguez}{2003}]{SchmiRodr03} Schmitt, B., \&  Rodriguez, J., 2003, GRL 108, 5104

\bibitem[\protect\citeauthoryear{Schutte and Khanna}{2003}]{SchuKha03} Schutte, W.~A. \& Khanna, R.~K., 2003, A\&A 98, 1049


\bibitem[\protect\citeauthoryear{Sofia et al.}{1994}]{Sofia_etal94} Sofia, U.~J., Cardelli, J.~A., \& Savage, B.~D., 1994, ApJ, 430, 650

\bibitem[\protect\citeauthoryear{Strazzulla et al.}{2001}]{Strazzulla_etal01} Strazzulla, G., Baratta, G.~A., Palumbo, M.~E. 2001, Spectrochimica Acta A 57, 825

\bibitem[\protect\citeauthoryear{Strazzulla}{2011}]{Strazzulla11} Strazzulla, G.,  2011, NIMB 269, 842

\bibitem[\protect\citeauthoryear{Tieftrunk et al.}{1994}]{Tieftrunk_etal94} Tieftrunk, A., Pineau des Forets, G., Schilke, P., \& Walmsley, C. M. 1994, A\&A, 289, 579

\bibitem[\protect\citeauthoryear{Vidal et al.}{2017}]{Vidal_etal17} Vidal, T.~H.~G., Loison, J.-C.,  Jaziri, Y.~A., Ruaud, M., Gratier, P., \& Wakelam, V. 2017, MNRAS, 469, 435

\bibitem[\protect\citeauthoryear{Viti et al.}{2004}]{Viti_etal04} Viti, S., Collings, M.~P., Dever, J.~W., McCoustra, M.~R.~S., Williams, D.~A. 2004, MNRAS, 354, 1141

\bibitem[\protect\citeauthoryear{Wakelam et al.}{2011}] {Wakelam_etal11} Wakelam, V., Hersant, F., Herpin, F. 2011, A\&A, 529, A112

    \bibitem[\protect\citeauthoryear{Woods et al.}{2015}]{Woods_etal15} Woods, P. ~M., Occhiogrosso, A., Viti, S., Ka\v{n}uchov\'{a},  Z., Palumbo, M.~E., \& Price, S.~D. 2015, MNRAS, 450, 1256

{ \bibitem[\protect\citeauthoryear{Zhang and Ewing}{2002}] {Zhang_Ewing02} Zhang, Z., Ewing, G.~E. 2002, Spectrochim. Acta 58, 2105}

{ \bibitem[\protect\citeauthoryear{Ziegler et al.}{2008}]{Ziegler_etal08}
 Ziegler J. F., Biersack, J. P., Ziegler M. D. 2008, The stopping and range
 of ions in solids (New York: Pergamon Press)}

\bibitem[\protect\citeauthoryear{Zolotov and Fegley}{1999}]{ZoloFeg99} Zolotov, M.J., \&  Fegley, Jr., B., 1999, Icarus 141, 40


\end{thebibliography}
\end{document}